\documentclass[sigconf]{acmart}

\usepackage{algorithmic}
\usepackage{graphicx}
\usepackage{textcomp}
\usepackage[most]{tcolorbox}
\usepackage{booktabs}
\usepackage{tabularx}
\usepackage{makecell}
\usepackage{subcaption}

\usepackage{tikz}
\usetikzlibrary{arrows.meta}
\usetikzlibrary{positioning}
\usetikzlibrary{fit,backgrounds,calc}

\AtBeginDocument{%
  }

\setcopyright{acmcopyright} 
\acmConference[EASE '26]{The 30th International Conference on Evaluation and Assessment in Software Engineering}{9--12 June 2026}{Glasgow, Scotland, United Kingdom}
\acmYear{2026}
\copyrightyear{2026}
\acmDOI{10.1145/XXXXXXX.XXXXXXX} 

\begin{document}




\title{Key Developer Roles and Organizational Coupling in Microservices: A Longitudinal Analysis}

\author{Xiaozhou Li}
\affiliation{%
  \institution{Free University of Bozen-Bolzano}
  \city{Bolzano}
  \country{Italy}}
\email{xiaozhou.li@unibz.it}

\author{Nariman Mani}
\affiliation{%
  \institution{Nutrosal Inc.}
  \city{Ottawa}
  \state{ON}
  \country{Canada}}
\email{nariman@nutrosal.com}

\author{Jose Sosa Rodriguez}
\affiliation{%
  \institution{University of Arizona}
  \city{Tucson}
  \state{AZ}
  \country{USA}}
\email{josesosa@arizona.edu}

\author{Tomas Cerny}
\affiliation{%
  \institution{University of Arizona}
  \city{Tucson}
  \state{AZ}
  \country{USA}}
\email{tcerny@arizona.edu}


\renewcommand{\shortauthors}{ et al.}

\begin{abstract}

Microservice-based systems impose significant organizational coordination challenges, yet the role of individual developers in shaping organizational coupling (OC) remains underexplored.
Prior work largely focuses on structural architectural aspects, leaving gaps in understanding how developer roles influence coordination dynamics over time. This study investigates how different developer roles contribute to OC in a large-scale microservices system. The analysis focuses on three key roles, namely Jacks, representing broad knowledge holders, Mavens, representing deep specialists, and Connectors, representing organizational bridges. A longitudinal repository mining analysis of GitHub data, including commits and issue and pull request interactions, is conducted to operationalize OC and quantify its evolution over time. The results show that Connectors are consistently associated with higher levels of OC, while the co-occurrence of multiple roles within the same developer further amplifies coupling effects. In contrast, Jacks and Mavens exhibit more localized and role-specific influences. These findings indicate that OC in microservices is primarily a role-driven phenomenon rather than an inevitable structural property, providing a foundation for role-aware organizational design and targeted decoupling strategies.

\end{abstract}




\maketitle

\section{Introduction}

Microservice architectures are widely adopted in the expectation that they enable faster delivery through independent deployment, localized change, and clearer service ownership \cite{dragoni2017microservices}. In this narrative, services define stable technical boundaries and teams align to those boundaries to reduce cross-team coordination and support Conway’s Law-style team architecture alignment. In practice, however, microservice ecosystems often exhibit persistent cross-service development activity, where a subset of developers contribute routinely across multiple services \cite{amoroso2023one}. This behavior creates coordination dependencies that undermine the intended organizational decoupling, even when services remain technically separable.

This study examines the phenomenon through the lens of \emph{organizational coupling (OC)}, defined as the interdependence among microservice teams that arises from shared developer activity across service boundaries \cite{li2023evaluating,li2025exploring}.
OC is measured from repeated cross-service contribution behavior in commits and issue and PR traces. Unlike technical coupling, which captures structural dependencies between services such as calls or shared interfaces, OC reflects the coordination load and ownership blur arising from developers' repeated cross-service work.
High OC can increase communication and review overhead, concentrate integration effort, and complicate socio-technical alignment, yet it is not an inherent property of microservices, but emerges from how work and knowledge are distributed over time \cite{li2023evaluating}.

Previous empirical work has characterized organizational implications of microservices and DevOps adoption and has emphasized the broader co-evolution of organizational and architectural structures \cite{tamburri2022orgarch,zhou2022ethnographic}. Meanwhile, research on key contributors has shown that a small fraction of developers often focus on critical knowledge and activity, raising concerns about resilience and sensitivity to change \cite{jabrayilzade2022busfactor}. Existing studies on microservice coupling also predominantly focus on technical dependencies, commit co-changes, or socio-technical congruence between architecture and communication structures \cite{panichella2021structural,driessen2023quantitative,zhong2024refactoring,herbsleb2016building}. Although informative, these approaches do not explicitly capture how organizational boundaries erode through developers' sustained cross-service contribution behavior. However, existing evidence provides limited explanatory power on a central question for microservice governance: \emph{who} primarily drives OC across services and \emph{how} the distribution of such drivers changes as the ecosystem evolves. 

The goal of this paper is to explain which key developer roles primarily drive OC in microservice ecosystems, how strongly these roles are associated with coupling intensity, and how changes in role distribution align with the evolution of coupling over time. To achieve this goal, a role-based perspective on key developers is adopted, focusing on three complementary roles, namely \emph{Jacks}, \emph{Mavens}, and \emph{Connectors}, which capture breadth of knowledge, depth of specialized expertise, and brokerage and coordination position, respectively \cite{ccetin2022analyzing,li2025toward}.
This perspective enables an individual-centered explanation of OC dynamics by identifying not only the presence of coupling, but also which role profiles are most responsible for it and how their distribution evolves over time. To operationalize this goal, the following research questions are investigated:

\begin{tcolorbox}[
    enhanced,
    colback=gray!5,
    colframe=black!70,
    left=2mm, right=2mm,
    overlay={
        \node[fill=black!70, text=white, font=\bfseries, anchor=west]
        at ([xshift=6pt,yshift=-8pt]frame.north west) {Research Question};
    }
]
\vspace{4mm}

\textbf{RQ1.} How do different key developer roles (Jack, Maven, Connector) differ in their contribution to organizational coupling?

\vspace{2mm}

\textbf{RQ2.} Which key developer roles are most strongly associated with high organizational coupling intensity?

\vspace{2mm}

\textbf{RQ3.} How does the distribution of key developer roles influence the evolution of organizational coupling over time?

\end{tcolorbox}

This paper presents a longitudinal repository mining study of a large-scale open-source system that analyzes traces of developer activity over time to quantify OC and its relationship with developer roles. Role metrics are computed over aligned sliding time windows to capture evolving knowledge and collaboration patterns, and these role distributions are related to service-level coupling intensity. This design supports a lifecycle-oriented view in which coupling can increase, stabilize, or decline under sustained development activity, depending on how cross-service work concentrates on specific role profiles.

The contribution of this paper is threefold: 1) Provides a longitudinal, role-centered empirical analysis of OC in a large microservice ecosystem by operationalizing Jack, Maven, and Connector roles within a unified artifact-traceability framework and relating them to coupling measurements; 2) Introduces and applies the Role Stacking Index (RSI) to quantify the co-occurrence of multiple key developer roles and analyze how multi-role concentration relates to coupling intensity; 3) Offers lifecycle-oriented evidence showing how changes in role distribution align with the evolution of OC over time, informing targeted organizational interventions for managing coordination and decoupling in microservice systems.

The remainder of the paper is structured as follows. Section \ref{sec:related} reviews related work on socio-technical alignment, microservice organization, and key contributor concentration. Section \ref{sec:method} presents the study design. Section \ref{sec:case} describes the longitudinal repository mining study and reports the results. Section \ref{sec:disc} discusses implications and limitations. Section \ref{sec:threats} covers the threats to validity, while Section \ref{sec:conclusion} concludes the paper.


\section{Related Work}
\label{sec:related}

A central research stream relevant to this study concerns the socio-technical relationship between organizational structure and software architecture. Foundational work like Conway’s Law suggests that system design mirrors the communication structure of the organization, a relationship further elaborated by the mirroring hypothesis, which posits that organizational coordination patterns are reflected in system modularity and vice versa \cite{conway1968committees}. Subsequent research on socio-technical congruence and social software engineering has emphasized the importance of alignment between coordination needs and communication structures to achieve effective software development outcomes \cite{cataldo2008socio,herbsleb2016building,tamburri2022orgarch}. 
This line of research provides a foundation for interpreting cross-boundary work as an organizational phenomenon. 
However, they primarily operate at the team or structural level and do not explicitly capture how individual developer roles contribute to or shape OC over time.

In the context of microservices, recent empirical work has examined the organizational implications of architectural decomposition and the adoption of DevOps.
For example, Zhou et al. report a cross-company ethnographic study of software teams adopting DevOps and microservices, documenting organizational arrangements, as well as observed benefits and issues \cite{zhou2022ethnographic}. Complementing these organizational observations, Zhong et al.\ provide evidence on how Domain-Driven Design (DDD) is applied to microservices as a boundary-design practice \cite{zhong2024ddd}. 
Although these studies provide valuable insights into how organizational and architectural boundaries are established and managed, they do not directly quantify OC between services from longitudinal development traces, nor do they attribute coupling dynamics to measurable developer roles.

A complementary line of work focuses on quantifying coupling and coordination from repository data. Previous studies have proposed to operationalize OC between microservices using contribution balance and switching behavior derived from repository traces and summarize coupling at the service-level \cite{li2023evaluating,li2025exploring}. Furthermore, Mani et al. provide empirical evidence that increases in OC systematically precede architectural degradation signals, establishing itself as a leading indicator of architectural instability \cite{mani2026organizational}. These approaches demonstrate that OC can persist even in systems with stable architectural decomposition. However, they typically treat developers as a homogeneous population and do not distinguish between different role profiles that may differentially contribute to coupling dynamics.

Another directly relevant stream addresses the risks associated with knowledge concentration and resilience, often framed through turnover sensitivity and "bus factor" style assessments. Jabrayilzade et al. propose a multi-modal approach to estimating bus factor by incorporating multiple activity channels beyond version control
\cite{jabrayilzade2022busfactor}. 
Similarly, Hajari et al. study how code review can spread knowledge and propose reviewer recommendation strategies that balance expertise, workload, and turnover risk \cite{hajari2024codereviewturnover}. 
Although these studies highlight the importance of key contributors and knowledge concentration, they do not differentiate between distinct developer roles or analyze how such roles influence cross-service OC.

Another line of work models developer knowledge and coordination structure with artifact traceability–based approaches that provide a foundation for identifying different roles of developers from repository data. 
Previous work models developer knowledge and coordination structure using traceability graphs, enabling the identification of roles such as broad knowledge holders, deep specialists, and organizational brokers \cite{ccetin2022analyzing,li2025toward}. These approaches motivate a role-sensitive perspective on developer activity, but have not been systematically connected to OC metrics or used to explain how coupling evolves over time in microservice ecosystems.

In summary, previous work provides strong foundations on 1) organization–architecture relationships and socio-technical alignment, 2) organizational practices and boundary design in microservices, 3) quantitative coupling metrics derived from repository traces, and 4) role-oriented modeling of developer activity. However, these research streams remain largely disconnected when it comes to explaining OC as an empirically observable and evolving phenomenon. In particular, existing studies rarely link identifiable developer role profiles with service-level coupling measured from longitudinal development data.

This gap limits both theoretical and practical understanding. From a theoretical perspective, it remains unclear whether OC should be interpreted as a structural property of the system or as a consequence of role concentration among a small subset of developers. From a practical perspective, the lack of role-sensitive explanations constrains the design of targeted interventions, such as redistributing coordination responsibilities or reducing dependency on key individuals. Hence, this paper addresses this gap by providing a role-centered longitudinal analysis that connects developer role distributions to OC dynamics in microservice ecosystems.

\section{Method}
\label{sec:method}

This paper presents a longitudinal repository mining study of developer activity in a large-scale microservice-based system.
Structured development artifacts, including commits, issues, and pull requests, extracted from GitHub are analyzed to quantify organizational coupling (OC) and its evolution over time. The approach follows a quantitative empirical design in which developer interactions with services are modeled as time-aware relationships. Metrics are defined to capture OC and developer roles, and these metrics are used to systematically answer the research questions. The study is longitudinal in that it examines how coupling and role distributions change across time windows, enabling the analysis of temporal dynamics rather than static snapshots.

\subsection{Key Developer Types}
\label{sec:keydev}

In this study, the concept of key developers is adopted and operationalized as proposed in prior work on organizational knowledge distribution and artifact traceability graphs \cite{ccetin2022analyzing,li2025toward}. Rather than treating developer importance as a monolithic construct, three complementary key developer roles are distinguished, namely \textit{Jacks}, \textit{Mavens}, and \textit{Connectors}, each capturing a different organizational function within a software project. This role-based perspective reflects the observation that developers contribute value through distinct patterns of knowledge breadth, specialization, and coordination, which cannot be adequately captured by simple activity counts alone.

The identification approach is grounded in an artifact traceability graph, where nodes represent developers and software artifacts (e.g., commits, files, issues), and edges encode traceable relationships such as authorship, modification, and linkage. This representation enables us to infer developer knowledge and interaction patterns from observable development activities, while accounting for recency effects and indirect relationships among artifacts. Following and adapting the definitions in the previous works, the three key developers types are as follows.

\begin{itemize}
    \item \textit{Jacks (Broad Knowledge Holders)} 
    
    Jacks are developers with broad, system-wide knowledge, reflected in sustained contributions across diverse files and modules. Jacks are identified using file coverage, defined as the proportion of project files reachable from a developer within a bounded-distance artifact traceability graph. High file coverage indicates versatile understanding that spans multiple parts of the system.
    
    \item \textit{Mavens (Deep Specialists)}

    Mavens are deep specialists who hold concentrated and often non-redundant knowledge of narrowly scoped artifacts. Mavens are identified by first detecting rarely reachable files, defined as files reachable by very few developers, and then computing each developer’s share of such files. High mavenness indicates critical expertise whose loss would pose disproportionate risk to the project.
    
    \item \textit{Connectors (Organizational Bridges)}
    
     Connectors link otherwise weakly connected developer groups and facilitate coordination and knowledge flow. Connectors are identified using betweenness centrality computed on a developer projection graph derived from artifact traceability relations. Developers with high centrality occupy bridging positions and play a key role in cross-boundary integration.

\end{itemize}

By distinguishing Jacks, Mavens, and Connectors, our approach captures three orthogonal dimensions of developer centrality: knowledge breadth, knowledge depth, and organizational mediation. These roles are not mutually exclusive; a developer may simultaneously exhibit characteristics of multiple roles. This fine-grained identification of key developers provides the analytical foundation for subsequent analyses of OC, risk concentration, and socio-technical structure evolution.

These three roles are selected because they capture complementary and analytically distinct dimensions of developer importance that are particularly relevant to OC in microservice ecosystems, namely breadth of system knowledge, concentration of specialized expertise, and coordination brokerage across boundaries. They also follow prior role-oriented work on artifact traceability and developer knowledge distribution, which provides a validated basis for their operationalization \cite{ccetin2022analyzing,li2025toward}. These roles do not exhaust all possible forms of developer importance; rather, they provide a focused and interpretable basis for studying how differentiated contribution patterns relate to coupling dynamics.

\subsection{Metrics and Computation}
\label{sec:metrics}

To operationalize the identification of key developer roles defined in Section~\ref{sec:keydev}, a set of graph-based metrics is computed on an \emph{artifact traceability graph} that captures developer interactions with software artifacts over time. This section describes the construction of the traceability graph, the reachability-based metrics used to identify Jacks and Mavens, and the network centrality metric used to identify Connectors.

\subsubsection{Artifact Traceability Graph Construction}

The project history is modeled as an undirected artifact traceability graph $G = (V, E)$, where vertices, hereafter referred to as nodes, represent developers and development artifacts, including commits, source files, and issues.
Edges encode traceable relations such as developer--commit authorship, commit--file modification, and commit--issue linkage.
This unified representation enables the analysis of both direct and indirect relationships between developers and artifacts.

To account for knowledge decay, edges induced by commits are weighted using a recency-aware distance function. Let $r \in (0,1]$ denote the normalized recency of a commit within the observation window. The distance associated with edges introduced by that commit is defined as $d = 1/r$. This formulation assigns shorter distances to more recent activities and longer distances to older interactions, while preserving the relative structure of the graph. All metrics are computed over sliding temporal windows of fixed length to ensure comparability over time and to prevent unbounded graph growth~\cite{ccetin2022analyzing}.

\subsubsection{Developer-Artifact Reachability}

Reachability is the core primitive used to infer developer knowledge. A file is considered \emph{reachable} by a developer if there exists a path from the developer node to the file node whose cumulative distance does not exceed a threshold $\theta$. 
To prevent artificial knowledge propagation, paths are not allowed to traverse intermediate developer nodes. This restriction limits artificial propagation of inferred knowledge and avoids conflating unrelated organizational boundaries. 

Let $R(d)$ denote the set of files reachable from developer $d$ under threshold $\theta$. The threshold is selected to exclude the oldest interactions while retaining the majority of recent development activity; following prior validation, $\theta$ is fixed to approximately cover the most recent $90\%$ of
the observation window.

\subsubsection{Metrics for Jack and Maven Identification}

\paragraph{File Coverage (Jack Metric)}
To quantify broad system knowledge, \emph{file coverage} is computed for each developer $d$ as follows:

\begin{equation}
\mathrm{Coverage}(d) = \frac{|R(d)|}{|F|}
\end{equation}

where $F$ is the set of all files present in the project during the observation
window. Developers with high coverage values are classified as Jacks,
reflecting familiarity across multiple modules and architectural layers.

\paragraph{Rare File Identification (Maven Metric)}
To capture specialization, \emph{rarely reachable files} are identified, defined as files that are reachable by at most $k$ developers, with $k = 1$ by default. Let $R_f$ denote the set of developers who can reach file $f$. A file is considered rare if $|R_f| \leq k$.

A developer's \emph{mavenness} score is computed as:

\begin{equation}
\mathrm{Mavenness}(d) =
\frac{|\{ f \in R(d) \mid |R_f| \leq k \}|}
     {|\{ f \mid |R_f| \leq k \}|}
\end{equation}

This metric captures the proportion of non-redundant knowledge held by a developer, highlighting expertise that is difficult to replace.

\subsubsection{Developer Projection and Connector Metric}

To identify Connectors, a \emph{developer projection graph} is derived from the artifact traceability graph. In this projection, nodes represent developers, and weighted edges reflect indirect relatedness through shared artifacts.

For each developer pair, all simple paths up to a bounded depth of four hops are enumerated in the artifact graph.
Recency weights are omitted in this step to focus on structural relatedness. Let $D$ denote the set of path lengths between two developers. These paths are aggregated using the reciprocal of the sum of reciprocal distances (RSRD):

\begin{equation}
\mathrm{RSRD} =
\left( \sum_{d \in D} \frac{1}{d} \right)^{-1}
\end{equation}

Smaller RSRD values indicate stronger relationships.
On the resulting weighted developer graph, normalized \emph{betweenness centrality} is computed. Developers with high betweenness centrality lie on a large fraction of shortest paths between other developers and are classified as Connectors, reflecting their role in cross-team coordination and information flow.

\subsubsection{Summary}

The three metrics, i.e., file coverage, mavenness, and betweenness centrality, are computed over a common traceability graph, ensuring internal consistency. Together, they provide orthogonal perspectives on the importance of developers: breadth
of knowledge (Jacks), depth of specialization (Mavens), and organizational mediation (Connectors). These metrics form the quantitative foundation for the OC analyses presented in the following sections.

It is noted that the main analysis parameters are selected to balance interpretability, temporal responsiveness, and consistency with prior work. The reachability threshold $\theta$ follows prior validation in artifact-traceability-based analysis and is set to retain the majority of recent development activity while excluding the oldest interactions within each window. For rare-file identification, $k = 1$ is used by default to capture strictly non-redundant knowledge and to maintain a conservative definition of specialization. The developer-projection analysis bounds path depth at four hops to capture meaningful indirect relatedness without introducing overly diffuse structural associations. Longitudinal role and coupling metrics are computed using 365-day windows with a 180-day step to balance stability against sensitivity to temporal change in a long-lived project. Where applicable, these choices follow prior formulations; otherwise, they are selected to provide interpretable and comparable measurements over time. Sensitivity to parameter choices remains a limitation, and future replications should assess alternative settings.

\subsection{Role Stacking Index}
\label{sec:rsi}

In microservice ecosystems, individual developers rarely embody a single organizational role. Although prior work distinguishes \emph{Jacks}, \emph{Mavens}, and \emph{Connectors} as analytically separable key developer roles, empirical evidence from the Spinnaker repository mining study indicates that these roles frequently co-occur within the same individual. This phenomenon is referred to as \emph{role stacking}. Role stacking captures the extent to which a developer simultaneously concentrates technical effort within services, as a Jack, holds rare or specialized expertise, as a Maven, and bridges organizational or service boundaries, as a Connector.

To operationalize this concept, the \emph{Role Stacking Index (RSI)} is introduced as a composite metric that quantifies the degree to which multiple key developer roles are jointly exhibited by a single developer within a given time window.
RSI is motivated by the observation that developers exhibiting stacked roles disproportionately influence OC, as they contribute to internal team stability while simultaneously intensifying cross-service dependencies through breadth of knowledge, depth of expertise, and brokerage position. As such, RSI provides an individual-centered lens for explaining system-level OC effects.

Let $d$ denote a developer observed within a fixed analysis window. From the key developer identification procedure, each developer is assigned three normalized role scores:

\begin{itemize}
  \item $J_d \in [0,1]$: \emph{Jack score}, reflecting the breadth of technical coverage across services;
  \item $M_d \in [0,1]$: \emph{Maven score}, reflecting the concentration of rare or exclusive expertise;
  \item $C_d \in [0,1]$: \emph{Connector score}, reflecting the developer's brokerage position in the collaboration network.
\end{itemize}
All role scores are independently normalized to ensure comparability across roles and time windows.

The Role Stacking Index of developer $d$ is defined as the geometric mean of the three role scores:

\begin{equation}
\mathrm{RSI}_d = \left( J_d \cdot M_d \cdot C_d \right)^{\frac{1}{3}}.
\end{equation}

The geometric mean is chosen for three reasons. First, it penalizes unbalanced role profiles: a developer excelling in only one or two roles achieves a low RSI. Second, it preserves scale invariance and bounds the metric in the range $[0,1]$. Third, it reflects the conceptual assumption that stacked roles interact synergistically rather than additively.

An RSI value close to $1$ indicates a strongly stacked role profile, where a developer simultaneously exhibits broad technical reach, exclusive expertise, and cross-team brokerage. Conversely, RSI values near $0$ characterize specialists or peripheral contributors whose influence is limited to a single organizational dimension. RSI does not replace individual role metrics; instead, it complements them by explicitly capturing \emph{role co-occurrence}, which is empirically associated with heightened OC intensity.

In the result analysis phase, RSI is computed at the developer level and subsequently aggregated at the service and time-window levels (e.g., mean or upper-percentile RSI per service). These aggregated RSI values are then related to Average Organizational Coupling (AOC) to assess whether services dominated by highly stacked developers exhibit systematically higher coupling. This enables a role-centric explanation of OC dynamics that extends beyond contribution volume or network centrality alone.

\subsection{Organizational Coupling}
\label{sec:org-coupling}

Organizational coupling (OC) characterizes the degree of interdependence between microservice teams induced by shared developer activity. Unlike technical coupling, which reflects structural dependencies among services, OC captures coordination demands arising from developers who contribute across service boundaries. High OC indicates blurred service ownership and increased cross-team coordination, while low coupling suggests clearer organizational separation.

This notion directly complements the key developer roles introduced in Section~\ref{sec:keydev}. In particular, \emph{Connectors} and \emph{Jacks} are expected to increase OC by spanning multiple services, whereas \emph{Mavens} typically reinforce service boundaries through deep, localized expertise. Therefore, OC provides a system-level perspective on how different key developer roles shape socio-technical structure.

Following prior work on organizational metrics in microservice ecosystems~\cite{li2023evaluating}, OC is defined at the service-pair level. Let $M_a$ and $M_b$ denote two microservices, and let $D=\{d_1,\dots,d_p\}$ be the set of developers who contribute to both services within a given observation window.
For each developer $d_i$, let $C_{i,a}$ and $C_{i,b}$ denote the number of commits to $M_a$ and $M_b$, respectively.
The OC between $M_a$ and $M_b$ is defined as:

\begin{equation}
OC(M_a,M_b)=
\sum_{i=1}^{p}
\left(
\frac{2C_{i,a}C_{i,b}}{C_{i,a}+C_{i,b}}
\cdot SD_i(M_a,M_b)
\right),
\label{eq:oc}
\end{equation}

where $SD_i(M_a,M_b)\in[0,1]$ is the switch degree of developer $d_i$, capturing how frequently the developer alternates contributions between the two services. The harmonic mean term emphasizes balanced cross-service contributions, while the switch degree accounts for temporal interleaving.

Because $OC(M_a,M_b)$ is sensitive to project scale and activity volume, normalization is required to enable comparison across services and time. OC is normalized relative to its theoretical maximum by assuming perfect alternation behavior, that is, $SD_i = 1$, for all shared developers:

\begin{equation}
NOC(M_a,M_b)=
\frac{
\sum_{i=1}^{p}
\left(
\frac{2C_{i,a}C_{i,b}}{C_{i,a}+C_{i,b}}
\cdot SD_i(M_a,M_b)
\right)
}{
\sum_{i=1}^{p}
\left(
\frac{2C_{i,a}C_{i,b}}{C_{i,a}+C_{i,b}}
\right)
},
\label{eq:noc}
\end{equation}

yielding a normalized value $NOC(M_a,M_b)\in[0,1]$.

To obtain a service-level view, the \emph{Average Organizational Coupling} (AOC) is computed for each microservice $M_a$:

\begin{equation}
AOC(M_a)=
\frac{1}{N-1}
\sum_{b\neq a} NOC(M_a,M_b),
\label{eq:aoc}
\end{equation}

where $N$ is the total number of microservices. The AOC reflects how strongly a service is organizationally tied to the rest of the system.

In summary, OC operationalizes cross-team interdependence induced by developer behavior. When analyzed jointly with the key developer roles in Section~\ref{sec:keydev}, it enables a fine-grained examination of how individual contribution patterns scale into the organizational structure at the system-level.

\begin{figure}[!ht]
  \centering
  \includegraphics[width=\linewidth]{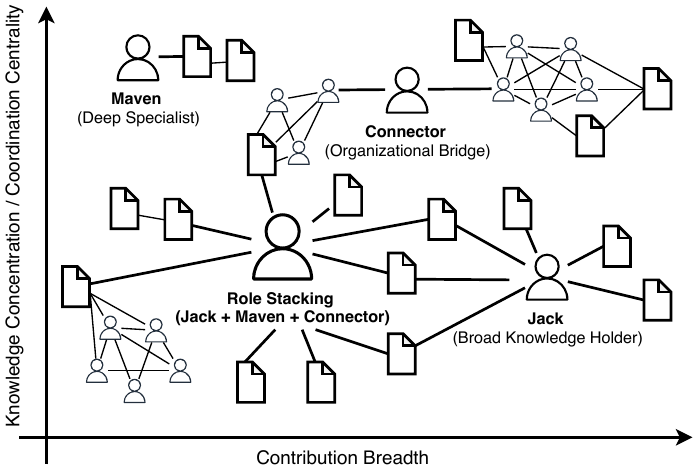}
    \caption{Conceptual model linking key developer roles and organizational coupling (OC), including the role-stacking dimension captured by RSI.}
  \label{fig:role-coupling-model}
\end{figure}

The role and coupling constructs introduced above are analytically related but capture different dimensions of socio-technical structure. Figure \ref{fig:role-coupling-model} summarizes the conceptual expectations that guide the analysis: broad cross-service contribution is expected to increase coupling differently from localized specialization, while brokerage positions are expected to intensify coupling through coordination across service boundaries. Role stacking provides a further individual-level lens by capturing when these dimensions co-occur within the same developers.

\subsection{Longitudinal Role-Coupling Analysis}
\label{sec:longitudinal}

This section describes the analytical procedure used to relate the temporal distribution of key developer roles to the evolution of OC, addressing RQ3. Building on the conceptual framing shown in Figure~\ref{fig:role-coupling-model}, the procedure examines whether the expected relationships between developer roles and coupling are reflected in the observed service-level coupling trajectories over time.

\paragraph{Step 1: Longitudinal observation windows.}
All role metrics (Jack coverage, Maven mavenness, Connector betweenness, and Role Stacking Index) and OC metrics are computed over aligned sliding time windows. This design ensures temporal comparability while capturing evolving contribution and coordination patterns without unbounded graph growth. Each window yields a snapshot of both role distribution and OC for every microservice.

\paragraph{Step 2: Service-level coupling trajectories.}
For each microservice, a time series of AOC values is constructed, representing the service’s organizational interdependence with the rest of the system over time. These AOC trajectories serve as the primary dependent variable for longitudinal analysis and enable observation of whether coupling increases, stabilizes, oscillates, or declines across successive windows.

\paragraph{Step 3: Temporal role distribution tracking.}
Within each time window, top-ranked Jacks, Mavens, and Connectors are identified per service, and aggregated role indicators, such as maximum role scores and aggregated RSI, are computed. By comparing these indicators across consecutive windows, a distinction is made between persistent role concentration, in which the same developers repeatedly occupy key roles, and role diffusion or turnover.

\paragraph{Step 4: Alignment of role persistence with coupling evolution.} 
Role distribution time series are aligned with AOC trajectories to examine how persistent versus transient role configurations correspond to changes in OC. This alignment reveals whether coupling growth coincides with sustained dominance of specific role profiles or with shifts in role occupancy over time.

\paragraph{Step 5: Analysis of Connector persistence and cross-service effects.}
Special attention is given to Connector roles, as these roles capture coordination brokerage across services. The analysis examines whether developers with persistently high betweenness centrality are associated with sustained or increasing AOC values, even during periods of stable architecture or stable commit volume. This step isolates coordination-mediated coupling effects from those attributable solely to contribution volume.

\paragraph{Step 6: Role stacking dynamics.}
Using the Role Stacking Index (RSI), the co-occurrence of multiple key roles within the same developers is analyzed over time. Changes in aggregated RSI values are compared with shifts in AOC to assess whether stacked role profiles amplify coupling trajectories beyond the effects observed for individual roles in isolation.

\paragraph{Step 7: Synthesis and interpretation.}
The observed temporal patterns are synthesized to derive a role-centric explanation of coupling evolution.
Persistent concentration of Connector roles and stacked role profiles is interpreted as a mechanism underlying cumulative and path-dependent OC, whereas role diffusion and turnover are associated with more elastic and adaptive coupling trajectories. This synthesis provides the explanatory basis for answering RQ3 in Section~4.2.4.

\subsection{Research Question Mapping}

The metrics introduced in this section support the research questions as follows. RQ1 is addressed by comparing how the role profiles of Jack, Maven and Connector contribute to OC patterns across services. RQ2 is addressed by relating role-specific metrics and aggregated RSI values to the service-level coupling intensity, measured through AOC. RQ3 is addressed through an aligned sliding-window analysis, which tracks how temporal changes in role distributions and RSI correspond to changes in OC over time.

The measures used in this study should be interpreted as operational proxies derived from repository traces rather than direct observations of formal responsibilities or organizational structure. In particular, Jack, Maven, and Connector roles reflect observable patterns of breadth, specialization, and brokerage in development activity, while OC captures an observable socio-technical proxy based on repeated cross-service work. These proxies are well suited to longitudinal repository-mining analysis, but they do not capture all organizational interactions, such as informal communication, managerial decisions, or undocumented coordination practices.

\section{Empirical Study and Results}
\label{sec:case}
This section examines a long-lived microservice ecosystem to identify developer role profiles, trace cross-service interaction over time, and characterize how role concentration aligns with observed coupling patterns.

\subsection{Data Collection and Analysis}
\label{sec:data}

A longitudinal repository mining study is conducted on Spinnaker, a large-scale open-source microservice-based system, using publicly available GitHub commit and issue data. Spinnaker provides an informative setting for this study because it is a large, long-lived open-source microservice system with sustained cross-service development activity, making it suitable for observing long-term role and coupling dynamics; however, it is not necessarily representative of all industrial microservice systems. Spinnaker consists of twelve core microservices that are developed in a polyrepo structure for most of its lifetime, providing a suitable context for analyzing developer roles and OC across service boundaries.

Two complementary data sources, namely commits and issues, are collected via the GitHub REST API. For each commit, the retrieved information includes the commit SHA, author identifier, timestamp, affected microservice, changed files, change type, including add, modify, delete and rename, and contribution size measured in lines of code (LOC). After excluding automated accounts, such as Dependabot and continuous integration bots, and resolving developer aliases, the final dataset comprises approximately 43{,}000 commits, 240{,}000 file-level change events, and 800 distinct human developers, covering the period from 2012 to April~2025. For issues and pull requests, collected data include metadata, such as creation and closure times and participant information, as well as full timeline events, including comments and commit references. Issue timelines are essential for establishing commit--issue links when constructing artifact traceability graphs. The resulting dataset contains approximately 2{,}000 issues and pull requests with 20{,}000 timeline events, involving around 600 developers. To mitigate identity fragmentation, multiple email aliases belonging to the same individual are manually and semi-automatically unified, a necessary preprocessing step for developer-centric analysis in large open-source software projects.

The commit and issue datasets are analyzed jointly. Commit data provide fine-grained evidence of cross-service contribution and switching behavior, while issue data enable artifact traceability and contextualize developer activity beyond code changes. Based on these data, two analytical products are derived, namely artifact traceability graphs for identifying key developer roles and contribution sequences across microservices for calculating OC. All analyses rely on descriptive statistics and network-based measures and are conducted over aligned temporal windows to support longitudinal analysis.

These preprocessing and extraction steps define the analysis pipeline used throughout the study and are designed to support the reproducibility of the reported measurements. The results are reported in terms of role distribution across services, the association between role profiles and coupling intensity, and the temporal alignment between role concentration and coupling trajectories.

\subsection{Results}
\label{sec:results}

The results are reported in three steps aligned with the research questions: role distribution and configurations across services (RQ1); role profiles associated with higher coupling using AOC and RSI (RQ2); and the alignment between role persistence and coupling trajectories over time through sliding-window analysis (RQ3).

\subsubsection{Role Configurations across Services}

Using the role metrics defined in Section~\ref{sec:method}, microservice-local \emph{Jacks}, \emph{Mavens}, and \emph{Connectors} are identified for each service and analysis window. The results show that these roles are not uniformly distributed across services. Instead, services differ in whether breadth of knowledge, specialization, and coordination brokerage are concentrated within the same individuals or distributed across multiple developers.

The first pattern is role stacking, where the same developer ranks highly across multiple role dimensions. For example, in services such as \texttt{clouddriver}, the top-ranked profile combines broad file reach, rare-file ownership, and strong brokerage. This indicates that the responsibility for the breadth, depth, and coordination is concentrated in a small number of individuals rather than distributed throughout the service.

\begin{table}[htbp]
\centering
\scriptsize
\setlength{\tabcolsep}{4pt}
\renewcommand{\arraystretch}{1.15}
\begin{tabularx}{\linewidth}{l X X X}
\toprule
\textbf{Service} & \textbf{Jack (Coverage)} & \textbf{Maven (Mavenness)} & \textbf{Connector (Centrality)}\\
\midrule

clouddriver & \makecell[l]{ezimanyi (0.228)\\ robzienert (0.163)\\ ajordens (0.143)
} & \makecell[l]{lwander (0.137)\\ Bijnagte (0.129)\\ ezimanyi (0.123)
} & \makecell[l]{ezimanyi (0.426)\\ cfieber (0.255)\\ dbyron-sf (0.252)
}\\
\hline



fiat & \makecell[l]{cfieber (0.252)\\ deverton (0.236)\\ ttomsu (0.230)
} & \makecell[l]{cfieber (0.267)\\ deverton (0.200)\\ jkschneider (0.133)
} & \makecell[l]{cfieber (0.318)\\ deverton (0.169)\\ ajordens (0.124)
}\\
\hline


gate & \makecell[l]{dbyron-sf (0.337)\\ ezimanyi (0.317)\\ ajordens (0.303)
} & \makecell[l]{dbyron-sf (0.270)\\ ezimanyi (0.163)\\ ajordens (0.128)
} & \makecell[l]{dbyron-sf (0.603)\\ ezimanyi (0.241)\\ ajordens (0.155)
}\\
\hline



kayenta & \makecell[l]{j-sandy (0.230)\\ fieldju (0.213)\\ marchello2000 (0.198)
} & \makecell[l]{j-sandy (0.259)\\ fieldju (0.182)\\ marchello2000 (0.140)
} & \makecell[l]{fieldju (0.338)\\ j-sandy (0.272)\\ marchello2000 (0.102)
}\\
\hline
keel & \makecell[l]{luispollo (0.204)\\ robfletcher (0.171)\\ emjburns (0.145)
} & \makecell[l]{luispollo (0.237)\\ emjburns (0.152)\\ robfletcher (0.111)
} & \makecell[l]{robfletcher (0.271)\\ luispollo (0.236)\\ emjburns (0.141)
}\\
\hline
orca & \makecell[l]{robzienert (0.190)\\ ajordens (0.175)\\ tomaslin (0.170)
} & \makecell[l]{Bijnagte (0.153)\\ robzienert (0.120)\\ ajordens (0.108)
} & \makecell[l]{robzienert (0.416)\\ marchello2000 (0.216)\\ dbyron-sf (0.194)
}\\


\bottomrule
\end{tabularx}
\caption{Representative role configurations in selected Spinnaker microservices (Full service-level rankings were computed for all services but are omitted for space).}
\label{tab:spinnaker_kd_top3_by_service}
\end{table}

In contrast, the second pattern is role separation, where the highest-ranked Maven is not the highest-ranked Connector, while broad contribution is also distributed differently. For example, \texttt{keel} and \texttt{orca} show partial separation between depth and brokerage, where the top Maven is not necessarily the top Connector. For such services, specialized knowledge and cross-service brokerage are less tightly fused, indicating a less concentrated coordination structure.

A third pattern is asymmetric concentration, in which one role is especially dominant. For example, some services show strong Connector prominence, suggesting that cross-service coordination is organized around a small number of brokers, whereas others are more strongly shaped by specialization and localized expertise. These differences indicate that services may face different organizational risks even when their total development activity is comparable.

Overall, these role configurations suggest that OC should be interpreted through \emph{who} contributes and \emph{how} they contribute, rather than through activity volume alone.

From a governance perspective, services with stacked or heavily brokered role configurations deserve closer attention because the knowledge and coordination load may be concentrated in a few developers. Therein, practical intervention can be adopted, e.g., mitigating Jack concentration motivates shared ownership of broad file surfaces; mitigating Maven concentration motivates redundancy on rare files and targeted documentation; mitigating Connector concentration motivates coordination load balancing and explicit boundary management (e.g., well-defined APIs and review rotation).

\subsubsection{Distinct Role Mechanisms of Organizational Coupling} Using service-level AOC together with the role metrics defined in Section~\ref{sec:method}, the contributions of Jack, Maven, and Connector roles to OC are compared through distinct behavioral mechanisms. The results show that these three roles differ not only in the strength of their association with coupling intensity, but also in the ways in which they generate or constrain cross-service interdependence.

Jacks exhibit broad cross-service contribution patterns, which are consistently associated with elevated OC. Services with strong Jack dominance, e.g., \texttt{clouddriver} and \texttt{orca}, tend to show the moderate-to-high AOC values, especially when the broad contribution is concentrated in a small number of developers. This pattern reflects that Jacks, by contributing to multiple services, create overlapping ownership boundaries that structurally entangle service teams. Importantly, Jack-driven coupling is primarily volumetric, as it emerges from frequent cross-service commits rather than explicit coordination. However, this effect is attenuated when breadth is shared among multiple contributors, suggesting that Jack-driven coupling is shaped by concentration and reach.

Mavens, identified through rare-file concentration and mavenness, show a different pattern. Although these developers hold critical localized expertise, services dominated by Maven roles, e.g., \texttt{fiat}, typically do not exhibit the highest AOC values. This suggests that specialization primarily reinforces service-local ownership rather than creating strong cross-service organizational entanglement. In Spinnaker, Maven roles rarely coincide with high cross-service recurrence, where Maven-dominated developers typically appear in only one or two services. As a result, Maven contributions tend to stabilize organizational boundaries rather than blur them, reinforcing clear ownership and reducing unnecessary developer overlap. In other words, Maven concentration is important for resilience and replaceability, but not the strongest standalone driver of coupling.

\begin{tcolorbox}[
    enhanced,
    colback=gray!5,
    colframe=black!70,
    left=2mm, right=2mm,
    overlay={
        \node[fill=black!70, text=white, font=\bfseries, anchor=west]
        at ([xshift=6pt,yshift=-8pt]frame.north west) {Answer to RQ1};
    }
]
\vspace{4mm}

Jack, Maven, and Connector roles contribute to OC through distinct mechanisms. Jacks increase coupling through broad cross-service contribution, Mavens mainly localize expertise within service boundaries, and Connectors most strongly amplify coupling by brokering coordination across services. Coupling is highest when these roles stack within the same developers, indicating that role concentration matters more than overall activity volume.

\end{tcolorbox}

Connectors, identified through betweenness centrality in the developer projection graph, show the strongest association with high OC. In Spinnaker, a small set of developers repeatedly appear as top-ranked Connectors in 8-10 out of the 12 services. Services with prominent Connectors, e.g., \texttt{gate} and \texttt{clouddriver}, repeatedly display higher AOC because these developers broker interaction across otherwise separate parts of the system. Unlike Jack-driven coupling, Connector-driven coupling is relational rather than volumetric. Even in services where Connectors are not the top code contributors, their mediation of issue discussions and cross-service coordination coincides with sharp increases in AOC, particularly during governance transitions and large-scale refactoring.

The strongest OC is observed when these role mechanisms stack within the same individuals.
Only a subset of services, notably \texttt{clouddriver} and \texttt{fiat}, exhibit high RSI values, where 
the same developers simultaneously exhibit breadth, specialization, and brokerage. These services also show the highest and most persistent AOC values over time. By contrast, services with low RSI (e.g., \texttt{gate}) maintain lower and more volatile coupling trajectories despite comparable development activity. This evidence indicates that OC increases non-linearly when breadth, specialization, and coordination brokerage converge in the same individuals.

\subsubsection{Key Developer Roles and OC Intensity}

To answer RQ2, role prominence is compared with service-level AOC by examining top role scores, the recurrence of the same developers across services, and aggregated RSI values. This comparison distinguishes roles that are merely present from roles that are consistently associated with high coupling intensity.

Connectors show the strongest association with high OC intensity. Services with consistently high AOC values, e.g., \texttt{clouddriver}, \texttt{orca}, and \texttt{gate}, are repeatedly characterized by prominent high-betweenness developers, and the recurrence of the same Connector profiles across multiple services indicates that brokerage is concentrated in a limited subset of individuals. Empirically, services where these developers act as primary Connectors exhibit AOC values frequently exceeding 0.25, whereas services with weak or fragmented Connector presence tend to remain below this level.  This pattern holds even when controlling for service size and commit volume, suggesting that Connector-driven coupling is not simply a byproduct of activity intensity. Instead, high Connector betweenness reflects persistent mediation of cross-service coordination, which directly amplifies OC.
 
Jack roles show a secondary but consistent association with coupling intensity. 
Services dominated by Jacks, identified via high file coverage ratios, e.g., \texttt{clouddriver}, generally exhibit medium to high AOC values. However, Jack effects are primarily service-local and do not consistently scale to system-wide coupling unless accompanied by Connector roles. Moreover, when Jack activity is distributed across multiple developers, the coupling intensity is attenuated. This indicates that Jack roles contribute to coupling primarily through contribution breadth, but their impact is bounded unless combined with coordination brokerage.

\begin{tcolorbox}[
    enhanced,
    colback=gray!5,
    colframe=black!70,
    left=2mm, right=2mm,
    overlay={
        \node[fill=black!70, text=white, font=\bfseries, anchor=west]
        at ([xshift=6pt,yshift=-8pt]frame.north west) {Answer to RQ2};
    }
]
\vspace{4mm}

Connector roles are most strongly associated with high organizational coupling intensity, because developers in brokerage positions repeatedly drive cross-service interdependence. Jack roles show a moderate association through broad contribution patterns, whereas Maven roles have weak standalone effects. The highest coupling intensity emerges when Connector roles stack with Jack and Maven characteristics in the same individuals.

\end{tcolorbox}

In contrast, Maven roles show the weakest standalone association with high OC intensity. Services with strong Maven concentration, e.g., \texttt{fiat} and \texttt{keel}, often exhibit relatively low AOC values, 
despite high internal specialization. Maven-dominated developers rarely recur across many services and seldom occupy high Connector positions. This suggests that mavenness primarily induces localized knowledge dependency rather than cross-service organizational entanglement. Although Maven concentration is critical from a resilience and bus-factor perspective, it does not drive high OC intensity by itself.

Meanwhile, aggregated RSI provides the clearest peak coupling intensity.
Services with high RSI values, e.g., \texttt{clouddriver} and \texttt{fiat} tend to be those where the same developers simultaneously contribute broadly, hold rare knowledge, and mediate coordination.
By contrast, services with low RSI, e.g., \texttt{kayenta} and \texttt{halyard}, 
maintain lower and more volatile AOC trajectories, even under sustained development activity. In this sense, RSI functions less as a separate role and more as an amplifier that reveals when multiple coupling-generating mechanisms have converged.
Practitioners should therefore monitor brokerage concentration and stacked role profiles in addition to contribution breadth.

\subsubsection{How Role Persistence Shapes Coupling Trajectories}

To answer RQ3, time-windowed role distributions are aligned with service-level AOC trajectories using the longitudinal procedure described in Section~\ref{sec:longitudinal}. This alignment enables comparison of whether coupling accumulates under persistent role concentration or remains more elastic when key roles diffuse or turn over across time.


\begin{figure}[!ht]
  \centering
  
  \begin{subfigure}{\linewidth}
    \centering
    \includegraphics[width=\linewidth]{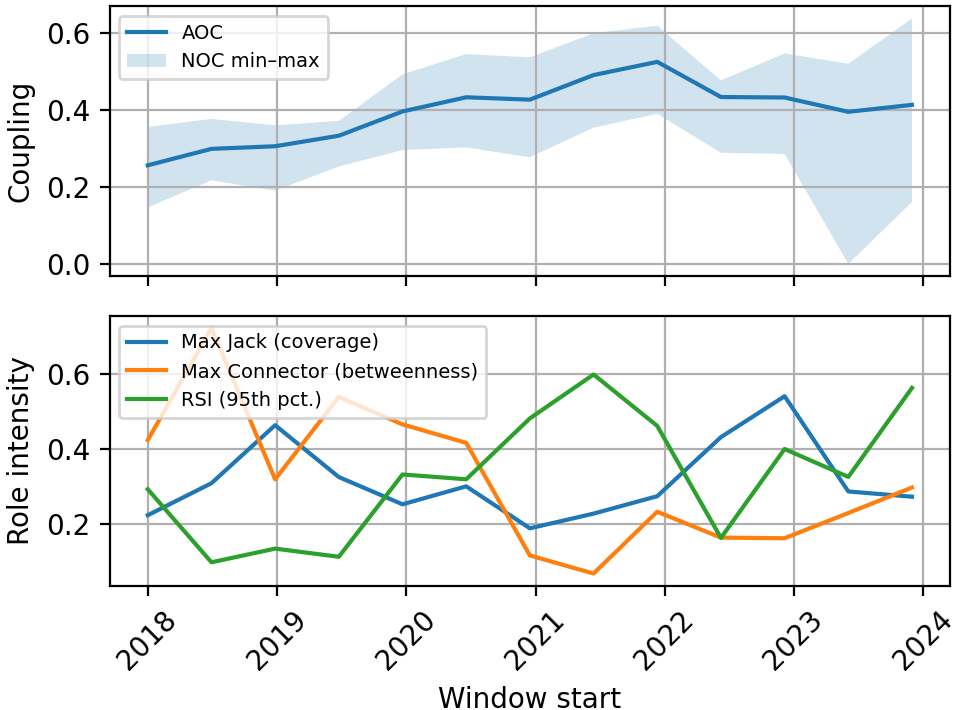}
    \caption{Service A (e.g., clouddriver): persistent role concentration and cumulative coupling.}
    \label{fig:temporal-role-coupling-a}
  \end{subfigure}
  
  \vspace{0.5em}
  
  \begin{subfigure}{\linewidth}
    \centering
    \includegraphics[width=\linewidth]{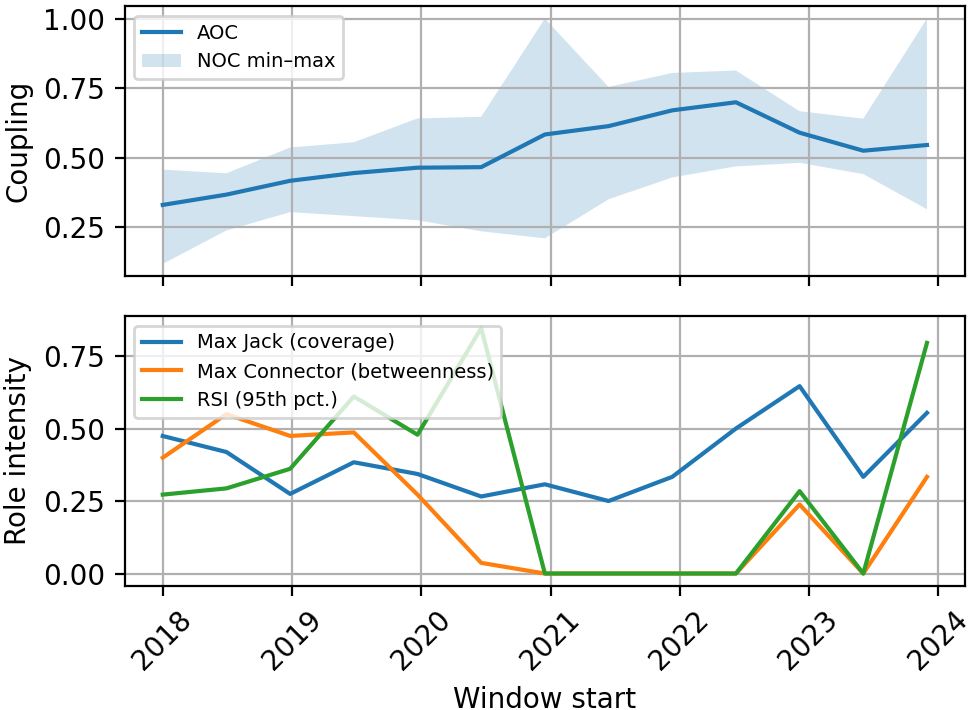}
    \caption{Service B (e.g., gate): role diffusion and elastic coupling.}
    \label{fig:temporal-role-coupling-b}
  \end{subfigure}
  
  \caption{Temporal alignment of organizational coupling and key developer role distributions for two representative Spinnaker services (365-day windows, 180-day step).}
  \label{fig:temporal-role-coupling}
\end{figure}

The \texttt{clouddriver} trajectory in Figure~\ref{fig:temporal-role-coupling} illustrates persistent concentration and cumulative coupling.
Across successive windows, the service maintains consistently high Connector betweenness and elevated RSI values, indicating sustained concentration of coordination brokerage and stacked role profiles. Correspondingly, AOC remains high and comparatively stable, suggesting that once cross-service coordination is routinized around a small number of developers, OC can harden rather than dissipate.

In contrast, the \texttt{gate} trajectory illustrates role diffusion and more elastic coupling.
Figure~\ref{fig:temporal-role-coupling} shows substantial variation in Jack and Connector dominance over time, along with consistently low RSI values. Despite sustained development activity and architectural centrality, the service maintains a relatively stable AOC below approximately 0.25, with a narrower NOC range. This indicates that the turnover or redistribution of key roles can prevent cross-service dependencies from becoming established.

Across both services, Connector persistence emerges as the clearest longitudinal signal of coupling growth. 
Periods in which Connector dominance is sustained coincide with elevated or increasing AOC, even when Jack coverage fluctuates. 
This suggests that sustained brokerage, especially when combined with stacked roles, amplifies coupling over time more strongly than broad contribution alone. Operationally, this means that persistent role concentration is a warning sign that OC may be becoming structural rather than temporary.


Across both trajectories, Connector persistence emerges as the clearest longitudinal signal of coupling growth, while increases in RSI tend to coincide with or precede upward shifts in AOC, particularly in \texttt{clouddriver}. This suggests that sustained brokerage, especially when combined with stacked roles, amplifies coupling over time more strongly than broad contribution alone. By contrast, services with consistently low RSI, such as \texttt{gate}, rarely exhibit sustained coupling growth.

\begin{tcolorbox}[
    enhanced,
    colback=gray!5,
    colframe=black!70,
    left=2mm, right=2mm,
    overlay={
        \node[fill=black!70, text=white, font=\bfseries, anchor=west]
        at ([xshift=6pt,yshift=-8pt]frame.north west) {Answer to RQ3};
    }
]
\vspace{4mm}


Organizational coupling evolves with changes in role distribution over time. Persistent concentration of Connector roles and stacked role profiles is associated with cumulative and rigid coupling trajectories, whereas role diffusion and turnover are associated with more elastic and adaptive coupling. This indicates that coupling is shaped by evolving role configurations, not by sustained development activity alone.

\end{tcolorbox}

\section{Discussion}
\label{sec:disc}

This study provides a role-centric interpretation of organizational coupling (OC) in microservice systems. The empirical results show that OC emerges from the level of developers' distribution of contribution breadth, specialization, and coordination brokerage, rather hand being solely an architectural decomposition consequence. In particular, Connectors and role stacking systematically shape the coupling intensity and evolution, indicating that cross-service interdependence is driven by persistent coordination behaviors.

Furthermore, the finding refines previous socio-technical perspectives, e.g., Conway’s Law, which emphasize the alignment between organizational structure and system architecture. Although these perspectives operate at the team or structural level, our results show that coupling can emerge even when architectural boundaries remain stable, driven by persistent cross-service activity concentrated in specific developer roles. This highlights a more dynamic, behavior-driven view of socio-technical alignment, where individual contribution patterns can reshape organizational dependencies independently of formal team structure.




The role-based view of OC has direct implications for the design and management of microservice teams. As Connector concentration is the strongest indicator of high OC, the systems should monitor them and distribute coordination responsibilities through clearer interfaces and shared workflows. Meanwhile, Jacks increase coupling through broad activity, but this effect weakens when contributions are shared. Promoting shared ownership can therefore reduce dependency without limiting knowledge diffusion. Furthermore, Maven roles primarily create knowledge concentration risks rather than coupling. Mitigation should focus on redundancy and knowledge transfer rather than structural changes. Importantly, 
high RSI indicates convergence of multiple coupling mechanisms, and such services should be treated as organizational hot-spots where responsibilities are redistributed. Overall, managing OC requires controlling how the responsibilities are distributed among developers, not just how the services are decomposed.




This study examines 
Spinnaker, which represents a complex and long-lived microservice ecosystem, but may not fully reflect all industrial settings.
It is treated as a critical example that reveals underlying mechanisms rather than universal effect sizes.
The identified mechanisms, particularly Connector persistence and role stacking, are likely transferable to systems with cross-service development activity, although their magnitude may vary depending on team structure, ownership practices, and repository organization. Future work should move beyond observational analysis towards intervention-oriented studies that evaluate how redistributing roles affects coupling and system evolution.

\section{Threats to Validity}
\label{sec:threats}

Threats to validity are discussed following the taxonomy proposed by Wohlin et al.~\cite{wohlin2012experimentation}.

Regarding internal validity, a potential threat arises from the fact that OC and developer roles are inferred from observational repository data, which does not permit direct control over alternative explanations such as organizational restructuring, changes in project governance, or external events, for example major releases or corporate transitions. To mitigate this threat, a longitudinal design with aligned time windows is adopted, enabling the distinction between persistent patterns and short-term fluctuations.
The consistent temporal alignment observed between role concentration and coupling trajectories strengthens the plausibility of the reported relationships.

For external validity, as mentioned above,  the study focuses on a single system. Although the chosen system provides a strong empirical basis, it is necessary to conduct further replication across diverse microservice ecosystems to assess generality. Future work should examine systems with different architectural styles, team structures, and organizational constraints to validate and refine the identified role-based mechanisms.

A potential threat to construct validity arises from the fact that developer roles such as Jack, Maven, and Connector are inferred from observed contribution patterns rather than directly measured responsibilities or self-reported expertise, which may lead to misclassification. Although these constructs are grounded in prior work on artifact traceability and OC, they serve as proxies for underlying socio-technical phenomena. In particular, informal coordination channels, such as meetings or chat-based communication, are not captured by the available data. This threat is addressed by explicitly interpreting the proposed measures as behavioral indicators of role enactment rather than as complete representations of organizational structure.

Regarding conclusion validity, threats include the sensitivity of results to parameter choices (e.g., threshold settings) and potential noise in data, e.g., automated commits or inconsistent author identities. These threats are addressed through systematic preprocessing steps, including identity consolidation, bot filtering, and normalized metrics. Furthermore, trends are interpreted at an aggregate and longitudinal level rather than relying on isolated data points, reducing the likelihood of spurious conclusions driven by short-term anomalies. Although statistical inference is not the main goal of the study, the consistency of observed patterns across services and over time supports the stability of the conclusions. 


\section{Conclusion}
\label{sec:conclusion}
This paper demonstrates that organizational coupling (OC) in microservice ecosystems is fundamentally a role-driven phenomenon rather than an inevitable consequence of architectural structure or development scale. Through a longitudinal repository mining study, qualitatively distinct influences of key developer roles on cross-service OC are identified. In particular, Connectors emerge as the primary drivers of high coupling intensity through sustained coordination brokerage, while Jacks contribute more moderately through broad cross-service activity, and Mavens largely stabilize service boundaries through localized expertise.
The strongest and most persistent coupling arises when these roles stack within the same individuals, thereby amplifying multiple coupling mechanisms simultaneously. The temporal analysis further shows that OC evolves in response to shifts in role distribution, with persistent role concentration leading to cumulative coupling, whereas role diffusion enables more elastic and adaptive coordination. Collectively, these findings refine socio-technical interpretations of Conway’s Law by showing that OC is not fixed by microservice decomposition alone, but is actively shaped by how key developer roles are distributed and enacted over time. This role-centric perspective provides both an explanatory lens for coupling dynamics and a foundation for targeted organizational interventions that support sustained microservice evolution.

\section*{Acknowledgment}

This material is based upon work supported by the National Science Foundation under Grant No. 2409933. Any opinions, findings, conclusions or recommendations expressed in this material are those of the author(s) and do not necessarily reflect the views of the National Science Foundation.


\bibliographystyle{ACM-Reference-Format}
\bibliography{bib}

\end{document}